\documentclass[letterpaper,english,aps,preprintnumbers,nofootinbib,showpacs,prd,notitlepage]{revtex4-1}

\usepackage{amsmath}
\usepackage{graphicx}
\usepackage{amsfonts}

\usepackage{slashed}

\begin{document}

\global\long\def\M{\mathcal{M}}
\global\long\def\pggg{\text{p-Ps}\to 3\gamma}
\global\long\def\pgg{\text{p-Ps}\to 2\gamma}
\global\long\def\pPs{\text{p-Ps}}
\global\long\def\piggg{\pi^{0}\to3\gamma}
\global\long\def\oggg{\text{o-Ps}\to3\gamma}
\def\resultWZ{1.1\cdot 10^{-80} \text{ eV}}

\title{Parapositronium decay into three photons and implications for the neutral pion}

\author{Andrzej Czarnecki}
\email{andrzejc@ualberta.ca}
\author{Divyesh Dagia}
\email{dagia@ualberta.ca}
\author{Ting Gao}
\email{tgao7@ualberta.ca}
\author{Ripanjeet Toor}
\email{ripanjee@ualberta.ca}

\affiliation{Department of Physics, University of Alberta, Edmonton, Alberta, Canada T6G 2E1}

\begin{abstract}
We complete the determination of the parapositronium decay into
three photons by evaluating amplitudes mediated by the $Z$ boson.
We show that, contrary to the expectation that the extra mass scale
$m_e$ may bring an enhancement to the overall scaling, the amplitude
turns out to start at $1/m_Z^6$ order, similarly to the $W$ boson
mediated amplitude. The decay rate with both $W$ and $Z$ boson
contributions is found to be $\resultWZ$, about 47 orders of
magnitude smaller than previously estimated. We also discuss its
implication for the $\pi^0\to3\gamma$ amplitude.
\end{abstract}
\maketitle

\section{Introduction}
In the ground state of positronium, the spins of the electron and the
positron add up to zero. This is the so-called para-positronium
(p-Ps), as opposed to the spin-one ortho-positronium (o-Ps), the
lowest excited state of Ps.

Because of charge conjugation symmetry ($C$-symmetry), within pure
quantum electrodynamics (QED), p-Ps decays only into an even
number
of photons, and o-Ps only into an odd number. This is important because
a decay of o-Ps must produce at least three photons and such a process
is about 1000 times slower than the leading two-photon decay of
p-Ps. The resulting longevity of o-Ps has many practical applications
\cite{charlton2001positron}. 

Does this  simple photon number parity rule hold beyond
pure QED, when $C$-violating weak interactions are included?
Landau-Yang theorem \cite{Landau:1948kw,Yang:1950rg} says that a
spin-one system cannot decay into two photons due to angular momentum
conservation and Bose symmetry. Thus  o-Ps cannot decay into
two photons under very general
assumptions, valid also for weak interactions.

Conversely, can p-Ps decay into three photons? Ref.~\cite{Bernreuther:1981ah} argued that yes
 and estimated the branching ratio of this
decay channel. This statement has been
supported in \cite{Pokraka:2017ore} by  an explicit calculation of
a gauge-invariant part of the amplitude, involving $W^\pm$ bosons. It
was found that the resulting decay rate is about 50 orders of
magnitude smaller than that estimated in  \cite{Bernreuther:1981ah}.

The  Standard-Model analysis \cite{Pokraka:2017ore} of the  $C$-violating channel $\pggg$ continues to be useful in several  contexts. On the high-energy side, the recent  surveys of few-body decays of  heavy Standard-Model particles use $\pggg$ as a template to estimate the rate of the analogous Higgs boson decay \cite{dEnterria:2312.11211,dEnterria:2508.00466}. The J-PET collaboration's papers on positronium decays reference it in their discussions of the $C$ violation relevant for symmetry tests \cite{Czerwinski:PoSDISCRETE2024,Czerwinski:2024,Eliyan:2024}. It is also incorporated into reviews of positronium precision physics \cite{Adkins2022,Bass:2019ibo}. Finally, it appears in the recent  work on positronium in external fields \cite{Kerbikov:2409.09496}. This interest motivates us to complete the Standard-Model treatment,  extending it from the $W$-mediated contribution to the full set including $Z$-mediated diagrams.

It was speculated in  \cite{Pokraka:2017ore} that the neutral $Z$ bosons may induce a
much more rapid decay than the charged $W^\pm$ due to the extra mass scale in the loop, the electron mass $m_e$; however, that second mass scale greatly complicates an explicit evaluation of the $Z$
boson loops. 
Here we compute the $Z$ boson effect  and thus complete the study of one-loop electroweak contributions to
$\pggg$. 

We find that adding the $Z$ effect  increases the branching ratio of the $\pggg$ by a factor of only about 50, in part due to the presence of a logarithmic enhancement by the factor $\ln (m_Z^2/m_e^2)\simeq 24$. However, the rate remains suppressed by twelve powers of the weak boson mass. While we have a reasonable understanding of this suppression in case of $W$ diagrams, as discussed \cite{Pokraka:2017ore}, its presence in the two-mass-scale $Z$ diagrams remains to be fully elucidated.

\section{Decay amplitude \label{sec:TotalAmp}}

\def\szer{40mm}
\begin{figure}[htb]
  \centering
  \begin{tabular}[c]{c@{\hspace{5mm}}c@{\hspace{5mm}}c@{\hspace{5mm}}c}
  \includegraphics[width=\szer]{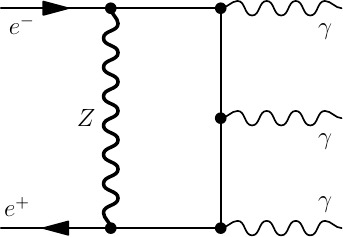}&
  \includegraphics[width=\szer]{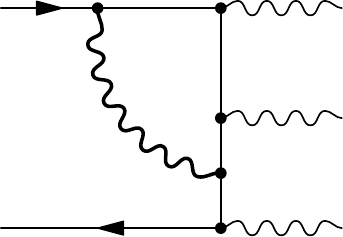}&
  \includegraphics[width=\szer]{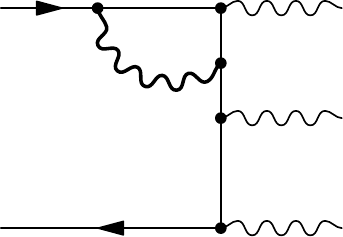} &
  \includegraphics[width=\szer]{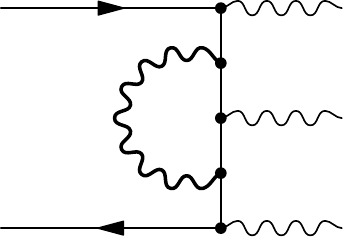}  
\\[1mm]
(a) & (b) & (c) & (d)  \\[4mm]
 \includegraphics[width=\szer]{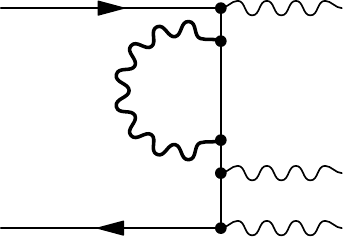}&
 \includegraphics[width=\szer]{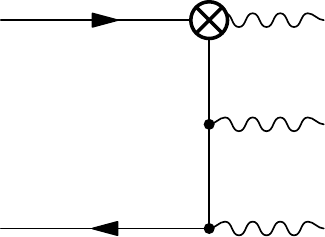}&
  \includegraphics[width=\szer]{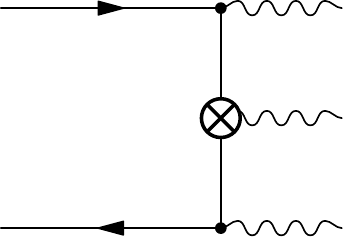} &
  \includegraphics[width=\szer]{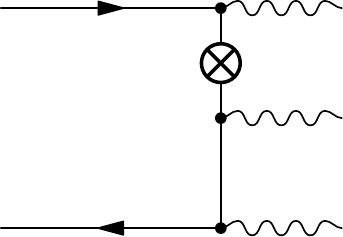}  
\\
(e) & (f) & (g) & (h)  
  \end{tabular}
  \caption{Contributions of the $Z$ boson (thick wavy lines) to the three-photon decay
    amplitude of p-Ps. Each diagram has six permutations of the photons. For diagrams (b,c,e), there are also diagrams with the $Z$ boson appearing at  locations symmetric to ones shown. Diagrams (c,d,e) are divergent and have  corresponding counterterm diagrams (f,g,h).}
  \label{fig:Zdiags}
\end{figure}
The $Z$-boson mediated $\pggg$ amplitude is generated by the diagrams shown in Fig.~\ref{fig:Zdiags}. In contrast to the $W$-boson mediated diagrams, when the $Z$-boson is replaced by a Goldstone boson, factors $\gamma^5$ are present in both vertices and do not induce the required $C$-violation. Therefore, there is no Goldstone contribution for the $Z$ diagrams and the gauge dependence has to cancel among diagrams (a-e) alone. Below we discuss the calculation of the amplitude in some detail.

We use natural units with unit speed of light and reduced Planck constant, $c=\hbar =1$.
\subsection{Projectors}
The amplitude can be written as a contraction of external spinors and photon polarization vectors with  factors which can only depend on the fermion and photon momenta,
\begin{equation}
    \mathcal{M}=\bar{v}(q_{+})\mathcal{M}^{\mu_1\mu_2\mu_3}(q_+,q_-,k_1,k_2,k_3)u(q_-)\epsilon_{\mu_1}^*(k_1)\epsilon_{\mu_2}^*(k_2)\epsilon_{\mu_3}^*(k_3).
\end{equation}
The spin of the electron and positron are anti-parallel in the parapositronium state, leading to the spin sum \cite{Czarnecki:1999mw},
\begin{equation}
    \bar{v}(q)\mathcal{M}^{\mu_1\mu_2\mu_3}u(q)\to\mathrm{Tr}\left[\frac{1}{\sqrt{2}}\frac{\slashed{q}+m_e}{2m_e}\gamma^5\mathcal{M}^{\mu_1\mu_2\mu_3}\right],
    \label{eq:2}
\end{equation}
where we use the observation that the spatial momenta of the electron and positron inside the positronium are small compared to its rest mass. We thus approximate their four-momenta by $q_{+}=q_{-}=q=(m_e,\boldsymbol{0})$.

On the photon side, one needs the parity ($P$)-violating combinations of three photon momenta, so that, when combined with the $P$-odd and $C$-odd weak vertex, the entire process is $P$-even and $C$-odd. 
This process is analogous to the decay of the neutral pion into three photons, analyzed by Dicus  \cite{Dicus:1975cz}. Here, as in  \cite{Pokraka:2017ore}, we follow his pioneering work. We note that his Eq.~(11) seems to have an extra overall factor of 3 which we have not been able to reproduce. We believe it is a typo; it propagates  to his  result for the rate in Eq.~(13). However, since his aim was only an estimate with an error bar spanning several orders of magnitude, this factor 3 is inconsequential. 

Contributions to our amplitude with three free indices satisfying the Ward identity and Bose symmetry can be decomposed into the following form \cite{Dicus:1975cz},
\begin{equation}
    \bar{v}(q)\mathcal{M}^{\mu_1\mu_2\mu_3}u(q)=F_1(x_1,x_2,x_3)T_1^{\mu_1\mu_2\mu_3}+F_2(x_1,x_2,x_3)T_2^{\mu_1\mu_2\mu_3}+F_3(x_1,x_2,x_3)T_3^{\mu_1\mu_2\mu_3}+F_4(x_1,x_2,x_3)T_4^{\mu_1\mu_2\mu_3},
\end{equation}
where
\begin{align}
    T_1^{\mu_1\mu_2\mu_3}&=\left(k_1^{\mu_3}-k_2^{\mu_3}\frac{k_1\cdot k_3}{k_2\cdot k_3}\right)\left(k_1^{\mu_2}k_2^{\mu_1}-g^{\mu_1\mu_2}k_1\cdot k_2\right),\\
    T_2^{\mu_1\mu_2\mu_3}&=\left(k_1^{\mu_2}-k_3^{\mu_2}\frac{k_1\cdot k_2}{k_3\cdot k_2}\right)\left(k_1^{\mu_3}k_3^{\mu_1}-g^{\mu_1\mu_3}k_1\cdot k_3\right),\\
    T_3^{\mu_1\mu_2\mu_3}&=\left(k_2^{\mu_1}-k_3^{\mu_1}\frac{k_2\cdot k_1}{k_3\cdot k_1}\right)\left(k_2^{\mu_3}k_3^{\mu_2}-g^{\mu_2\mu_3}k_2\cdot k_3\right),\\
    T_4^{\mu_1\mu_2\mu_3}&=k_1^{\mu_2}k_2^{\mu_3}k_3^{\mu_1}-k_1^{\mu_3}k_2^{\mu_1}k_3^{\mu_2}+g^{\mu_1\mu_2}\left(k_1^{\mu_3}k_2\cdot k_3-k_2^{\mu_3}k_1\cdot k_3\right)\nonumber\\
    &\qquad+g^{\mu_2\mu_3}\left(k_2^{\mu_1}k_3\cdot k_1-k_3^{\mu_1}k_2\cdot k_1\right)+g^{\mu_3\mu_1}\left(k_3^{\mu_2}k_1\cdot k_2-k_1^{\mu_2}k_3\cdot k_2\right),
\end{align}
\begin{equation}
    x_1=\frac{k_1\cdot q}{m_e^2}, \quad x_2=\frac{k_2\cdot q}{m_e^2}, \quad x_3=\frac{k_3\cdot q}{m_e^2},
\end{equation}
and we have used momentum conservation $k_1+k_2+k_3=2q$ to eliminate the scalar products $k_i\cdot k_j$ that may appear in the form factors $F_1,\cdots,F_4$.
The Levi-Civita tensor does not enter the possible combinations due to the wrong parity. Each of the four structures separately satisfy the Ward identity $k_{i\mu_i}T_n^{\mu_1\mu_2\mu_3}=0$, where $i=1,2,3$ and $n=1,2,3,4$. Further, the Bose symmetry leads to the relations
\begin{align}
    F_2(x_1,x_2,x_3)=F_1(x_1,x_3,x_2),\quad F_3(x_1,x_2,x_3)=F_1(x_2,x_3,x_1),
    \label{eq:F2F3}
\end{align}
so $F_1$ and $F_4$ are the only two form factors not related by symmetry. The projectors onto the corresponding form factors are given by
\begin{align}
    &P_1=-\frac{(k_2\cdot k_3)}{2(k_1\cdot k_2)^3(k_1\cdot k_3)}T_1-\frac{1}{4(k_1\cdot k_2)^2(k_1\cdot k_3)}T_4,\\
    &P_4=-\frac{1}{4(k_1\cdot k_2)^2(k_1\cdot k_3)}T_1+\frac{1}{4(k_1\cdot k_3)^2(k_1\cdot k_2)}T_2-\frac{1}{4(k_2\cdot k_3)^2(k_1\cdot k_2)}T_3-\frac{1}{2(k_1\cdot k_2)(k_1\cdot k_3)(k_2\cdot k_3)}T_4.
\end{align}
With the fermion spin sum and the projectors, the form factors are given by
\begin{align}
    &F_1=\mathrm{Tr}\left[\frac{1}{\sqrt{2}}\frac{\slashed{q}+m_e}{2m_e}\gamma^5\mathcal{M}^{\mu_1\mu_2\mu_3}\right]P_{1,\mu_1\mu_2\mu_3}\\
    &F_4=\mathrm{Tr}\left[\frac{1}{\sqrt{2}}\frac{\slashed{q}+m_e}{2m_e}\gamma^5\mathcal{M}^{\mu_1\mu_2\mu_3}\right]P_{4,\mu_1\mu_2\mu_3}.
\end{align}

\subsection{Renormalization of the $C$-violating amplitude}
\label{sec:renorm_axial}

In pure QED the decay $\text{p-Ps}\to 3\gamma$ is forbidden by charge
conjugation: the ground-state parapositronium has $C=+1$, while a
$3\gamma$ state has $C=(-1)^3=-1$. Hence the amplitude vanishes
identically in a $C$-invariant theory. In the Standard Model the weak
interactions violate $C$ (and $P$), and the leading nonzero contribution
arises from one-loop graphs containing $W$ or $Z$ bosons. These loops
generate an axial-vector structure on the electron line, the
necessary ingredient for a nonzero $C$-violating trace.

Although the decay amplitude starts at one loop, ultraviolet divergences
can still appear through divergent subgraphs (electron self-energy and
electron-photon vertex corrections). 
Renormalizability implies that such
divergences are removed by the counterterms of the underlying theory.
In the on-shell scheme this is  implemented by including
counterterm diagrams, shown in Fig.~\ref{fig:Zdiags}(f-h), while external-leg corrections
are absorbed into the field renormalization constants \cite{Denner:1991kt}.

Because the weak interactions are chiral, the electron field $\psi$
renormalizes with independent left- and right-handed constants \cite{Denner:1991kt},
\begin{equation}
\psi_{0L}=\sqrt{Z_L}\,\psi_L,
\qquad
\psi_{0R}=\sqrt{Z_R}\,\psi_R,
\end{equation}
where $Z_L$ differs from $Z_R$ once weak loops are included. In terms of the Dirac
field $\psi=\psi_L+\psi_R$ this can be rewritten as
\begin{equation}
\psi_0=\left(1+\frac12\,\delta Z_V+\frac12\,\delta Z_A\,\gamma_5\right)\psi,
\qquad
\delta Z_V=\frac12(\delta Z_L+\delta Z_R),\quad
\delta Z_A=\frac12(\delta Z_R-\delta Z_L).
\label{eq:ZVZA_def}
\end{equation}
The parameter $\delta Z_A$ expresses the parity-violating (axial) part of
the electron wave-function renormalization; it vanishes in any parity-symmetric limit where $Z_L=Z_R$.

Correspondingly, the renormalized electron self-energy admits the
decomposition
\begin{equation}
\Sigma(\slashed p)= \slashed p\,\Sigma_V(p^2)
+\slashed p\,\gamma_5\,\Sigma_A(p^2) + \ldots,
\end{equation}
and, isolating the axial contribution,
\begin{equation}
\slashed p\,\gamma_5\,\Sigma_A(p^2)=\slashed p\,\gamma_5\,
\Bigl(\widehat\Sigma_A(p^2)-\delta Z_A\Bigr),
\label{eq:self_axial}
\end{equation}
where $\widehat\Sigma_A$ denotes the loop contribution prior to adding
the counterterm. In an on-shell scheme one may impose that the axial part
of the inverse propagator vanishes at the physical pole,
which fixes
\begin{equation}
\delta Z_A=\widehat\Sigma_A(\slashed p = m_e)\,.
\label{eq:deltaZA_OS}
\end{equation}

Finally, expanding the bare electromagnetic interaction
$e_0\,\bar \psi_0\gamma^\mu \psi_0 A^0_\mu$ in terms of renormalized fields
shows that the chiral wave-function renormalization induces an axial
counterterm in the electron-photon vertex,
\begin{equation}
\Gamma^\mu_A = i e\,\delta Z_A\,\gamma^\mu\gamma_5\,.
\label{eq:vertex_axial_ct}
\end{equation}
Since the decay is $C$-violating,
only the axial ($\gamma^\mu\gamma_5$) structure contributes, so at this
order the renormalization of the amplitude is completely controlled by
$\delta Z_A$ through
Eqs.~\eqref{eq:self_axial}-\eqref{eq:vertex_axial_ct}.

\subsection{Expansion in regions}

In case of $Z$ boson contributions in Fig.~\ref{fig:Zdiags},  loop
integrals contain two mass scales, $m_e$ and $m_Z$. While it is
possible to reduce the result to the Passarino-Veltman
function~\cite{Passarino:1978jh}, for our purpose it is sufficient to
expand the result in powers of $m_e/m_Z$ to the lowest not-vanishing
order. We expand the loop integral in two regions \cite{AE}: the hard region where the loop momentum scales like $m_Z$, and the soft region where the loop momentum scales like $m_e$.

To count powers of $1/m_Z$, we assign the following scaling in each
region \cite{AE}:
\begin{align}
    \text{Soft region:}&\quad m_Z\sim m_Z/\lambda,\\
    \text{Hard region:}&\quad l^\mu\sim l^\mu/\lambda,\quad   m_Z\sim m_Z/\lambda,
\end{align}
where $\lambda$ is a small parameter, $l$ is the loop momentum, and the unlisted quantities do not scale with $\lambda$. In each region, the loop integral is expanded to $1/m_Z^6$ and then evaluated using the dimensional regularization with $D=4-2\epsilon$. After performing the integration-by-parts reduction, the loop integral is reduced to the following master integrals,
\begin{align}
    A_0(m^2)&=\int { \text{d}^D l \over (2\pi)^D } \frac{1}{l^2-m^2},\\
    B_0(p)&=\int { \text{d}^D l \over (2\pi)^D } \frac{1}{(l^2-m_e^2)\left[(l-p)^2-m_e^2\right]},\\
    C_0(p_1,p_2)&=\int { \text{d}^D l \over (2\pi)^D } \frac{1}{(l^2-m_e^2)\left[(l-p_1)^2-m_e^2\right]\left[(l-p_1-p_2)^2-m_e^2\right]},
\end{align}
which  result in
\begin{align}
    A_0(m^2)&=\frac{im^2}{16\pi^2}\left(\frac{1}{\epsilon}-\gamma_E+1-\ln(m^2)\right),\\
    B_0(p)&=\frac{i}{16\pi^2}\left(\frac{1}{\epsilon}-\gamma_E+2-\ln(m_e^2)-2\sqrt{\frac{4m_e^2-p^2}{p^2}}\arcsin\sqrt{\frac{p^2}{4m_e^2}}\right), \quad 0<\frac{p^2}{4m_e^2}<1,\\
    C_0(p_1,p_2)&=-\frac{i}{16\pi^2}\frac{1}{p_1\cdot p_2}\arcsin^2\sqrt{\frac{p_1\cdot p_2}{2m_e^2}},\quad 0<\frac{p_1\cdot p_2}{2m_e^2}<1.
\end{align}

\subsection{Results}

As a consistency check we work in the $R_\xi$ gauge. After including the counterterms, the $\xi$-dependence in the soft and hard region cancel separately, and when the two regions are summed together we arrive at the following finite results for $F_1$:
\begin{align}
    F_1(x_1,x_2,x_3)=& -\frac{\sqrt{2}m_ee^3 g^2 g_A g_V}{720\pi^2m_Z^6\cos^2\theta_W}(1-x_1)\left[f_1(x_1,x_2)-f_1(x_2,x_1)\right] \label{eq:F1} \\
    f_1(x_1,x_2)=&\frac{(x_1-x_2)\left\{917\left[(1-x_1)^2+(1-x_2)^2-1\right]+x_1x_2\left[1251-47(x_1+x_2)\right]\right\}}{4x_1x_2(1-x_1)(1-x_2)} \nonumber\\
    &+\frac{15(x_1-x_2)(1-x_1-x_2)}{x_1 x_2}\ln\frac{m_Z^2}{m_e^2}
\nonumber\\
    &+\frac{30(2x_1^2+2x_1x_2-8x_1-15x_2+15)}{x_1(1-x_1)}\sqrt{\frac{x_1}{1-x_1}}\arcsin\sqrt{1-x_1} \nonumber\\
    &+\frac{45(2x_1^2+4x_1x_2-6x_1-5x_2+5)}{x_1(1-x_1)^2}\arcsin^2\sqrt{1-x_1},
\end{align}
where $\theta_W$ is the weak mixing angle, $e=g\sin\theta_W$ is the charge of the positron, the vector and axial couplings of the $Z$-boson to the electron are $g_V=-1/2+2\sin^2\theta_W$, $g_A=-1/2$, and $F_2$ and $F_3$ are given by Eq.~\eqref{eq:F2F3}.
Remarkably, the non-vanishing contribution starts at the $1/m_Z^6$
order (we have not found a simple physical explanation, but the
cancellation  of less suppressed terms, like $1/m_Z^2,1/m_Z^4$,
is exact in our analytic result), and  the product  $\mathcal{M}\cdot T_4$ vanishes, so $F_4$ is also fully determined by $F_1$ and its permutations:
\begin{align}
    &F_4(x_1,x_2,x_3)=\frac{1-x_3}{2(1-x_1)}F_1(x_1,x_2,x_3)-\frac{1-x_2}{2(1-x_1)}F_2(x_1,x_2,x_3)+\frac{1-x_1}{2(1-x_2)}F_3(x_1,x_2,x_3).  \label{eq:F4}
\end{align}

\subsection{Vanishing diagrams} \label{sec: fermion loops}
\begin{figure}[htb]
  \centering
   \begin{tabular}[c]{c@{\hspace{5mm}}c@{\hspace{5mm}}c@{\hspace{5mm}}c}
  \includegraphics[width=\szer]{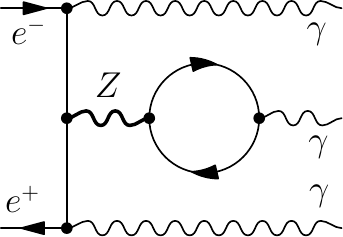}&
  \includegraphics[width=\szer]{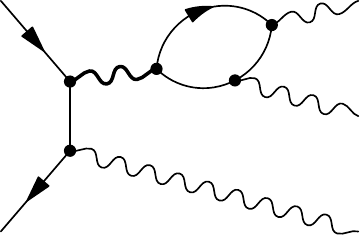}&
  \includegraphics[width=\szer]{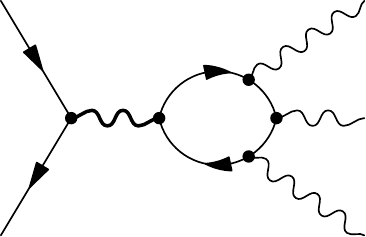} &
  \includegraphics[width=\szer]{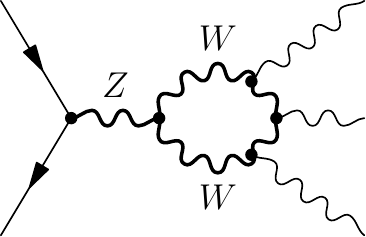}  
\\[1mm]
(a) & (b) & (c) & (d)  \\[4mm]
 \includegraphics[width=\szer]{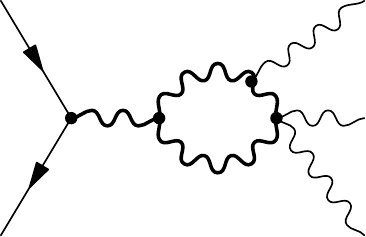}&
 \includegraphics[width=\szer]{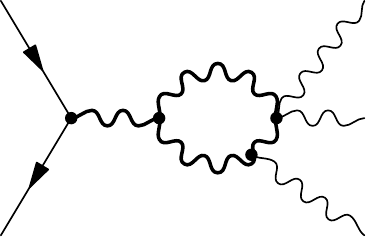}&
  \includegraphics[width=\szer]{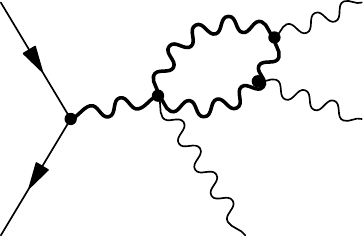} &
  \includegraphics[width=\szer]{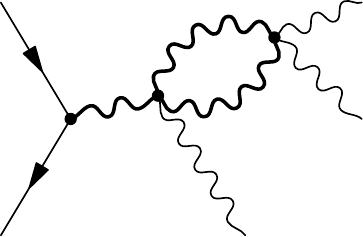}  
\\[1mm]
(e) & (f) & (g) & (h)  
  \end{tabular}
  \caption{Diagrams that do not contribute to the amplitude for  reasons explained in the text. Diagrams (a,b,c) contain a virtual fermion loop. In diagrams (d-h), the virtual loop has a $W$ boson. Both $Z$ and $W$ bosons are drawn with a thicker wavy line, while the photons are depicted with a regular wavy line. }
  \label{fig:fermionloops}
\end{figure}

Before closing this Section, we note that other diagrams, shown in Fig.~\ref{fig:fermionloops}, look like they may  contribute to $\pggg$ but in fact do not. In the following we present arguments that eliminate these diagrams.

The fermion loop in diagram~\ref{fig:fermionloops}(a), and its analog with $W$-boson replacing the fermion, induce the mixed $Z-\gamma$ vacuum polarization which in general can be written as
\begin{equation}
    \Pi^{\mu\nu}(k)=\Pi(k^2)\left(k^2 g^{\mu\nu}-k^\mu k^\nu\right).
\end{equation}
When the photon goes on-shell, $k^2$ vanishes  and the $k^\mu k^\nu$ term vanishes after contraction with the photon polarization vector. Therefore, this diagram does not contribute.

Diagram~\ref{fig:fermionloops}(b) was mentioned in \cite{Bernreuther:1981ah} as one of possible ways to generate the $\pggg$ decay. For this diagram, the $Z$-boson vertex has to couple axially on the $\pPs$ side and as a vector on the fermion loop side to generate the correct $C$ and $P$; however, a pure three-vector-boson coupling generated by the fermion loop is forbidden by Furry's theorem~\cite{Furry:1937zz}. As a consistency check, we also computed this diagram exactly to show that it indeed vanishes. For the same diagram with the $W$ loop, Furry's theorem does not apply, however, as shown in~\cite{Gounaris:2000tb}, the three neutral boson vertex cannot be generated by pure boson loops because the pure boson interactions are $CP$-conserving and the allowed $CP$-conserving tensor structures all carry $\epsilon^{\mu\nu\alpha\beta}$, which cannot be generated by pure boson interactions.

Diagram~\ref{fig:fermionloops}(c) is forbidden by gauge invariance:
the trace of the p-Ps spin projector (see Eq.~\eqref{eq:2}) with the
$Z$-electron vertex must be proportional to the total four-momentum
$k_{Z^\star}^\mu = 2q^\mu$ flowing in the $Z$-propagator, since there
is no other four-vector and the trace contains a single Lorentz index,
\begin{equation}
    \mathrm{Tr}\left[\frac{1}{\sqrt{2}}\frac{\slashed{q}+m_e}{2m_e}\gamma^5(\gamma^\mu\gamma^5) \right] = 
 -\frac{k_{Z^\star}^\mu}{\sqrt{2}m_e}    .
\end{equation}
We note that only the axial $Z$-electron vertex contributes to this trace because of the $\gamma^5$ in the p-Ps spin projector (a non-zero trace containing a single $\gamma^5$ requires four momenta or Lorentz indices). 

In order to violate $P$-symmetry, both the axial and the vector vertices of $Z$ must contribute, so the vertex on the virtual fermion loop must be the vector. But a correlator of four vector currents, contracted with the momentum of the external $Z$, vanishes by gauge invariance,
\begin{equation}
    k_{Z^\star}^\mu  \Gamma_{\mu \mu_1 \mu_2 \mu_3}(Z^\star \to 3\gamma)=0\quad \text{(vector Ward–Takahashi identity)}. 
\end{equation}

We have paid so much attention to the vanishing of this diagram because the estimate of $\Gamma(\pggg)$ in Ref.~\cite{Bernreuther:1981ah} was based on the power counting of parameters characterizing it. 

The sum of diagrams, shown in Fig.~\ref{fig:fermionloops}(d-h), in which the virtual $Z$ decays into three photons via $W$-boson loops vanishes  for the same reason: the $Z\gamma\gamma\gamma$ amplitude is transversal with respect to the four-momentum of $Z$, $k_{Z^\star}$
\cite{Glover:1993nv}.

Another way to see the vanishing of diagrams
\ref{fig:fermionloops}(c-h) is to note that there is no other diagram that can cancel the gauge-dependence in the $Z$ propagator, and again because the only available vector is $q^\mu$, the gauge-independent contribution and the gauge-dependent contribution are proportional to each other, so the only way to be consistent is for this diagram to vanish.

\section{Decay rate \label{sec:Decayrate}}

The amplitude we get for the $Z$-boson mediated diagrams interferes with the $W$-boson mediated diagrams computed in~\cite{Pokraka:2017ore}. To calculate the decay rate for $\pggg$, we include the amplitude for both sets of diagrams. The amplitude in Sec.~\ref{sec:TotalAmp} has the electron and the positron in the initial state. Because they form the bound $\pPs$ state, the integration over their phase space is weighted by the wave function. Integration over their relative motion reduces the decay rate to \cite{Peskin:1995ev}
\begin{equation}
\Gamma=\left\vert\sqrt{2M}\psi_\mu(r=0)\right\vert^2\Gamma_{\text{three-body}},
\end{equation}
where $\sqrt{2M}$ converts our non-relativistic normalization to the relativistic normalization, $M=2m_e$ is the mass of the $\pPs$ bound state, $\mu=m_e/2$ is the reduced mass and $\Gamma_{\text{three-body}}$ is the standard decay rate for three-body decay. Using the results in~\cite{ParticleDataGroup:2024cfk}, the decay rate is given by
\begin{equation}
    \Gamma(\pggg) = \left\vert\psi_\mu(r=0)\right\vert^2
    \times\frac{1}{3!}
    \frac{m_e^2}{32\pi^3}
    \int_0^1 dx_1\int_{1-x_1}^{1}dx_2
    \left\vert\mathcal{M}\right\vert^2,
\end{equation}
where the overall factor $1/3!$ accounts for identical particles in the final state and momentum conservation $x_1+x_2+x_3=2$ is used to eliminate $x_3$. 

It is convenient to change variables to $u=1-x_1$ and $z=1-x_2$ to parameterize the  kinematic invariants,
\begin{equation}
k_1\!\cdot\!k_2 = 2(1-u-z)\,m_e^2, \qquad
k_1\!\cdot\!k_3 = 2z\,m_e^2, \qquad
k_2\!\cdot\!k_3 = 2u\,m_e^2,
\label{eq:uzparam}
\end{equation}
The decay rate is
\begin{align}
\Gamma(\pggg) = {m_e^{11} \alpha^3 \over 96 \, \pi^4} &\int_0^1 \text{d} z \int_0^{1-z} \text{d}u { 1-u-z \over uz}
\,\left[ 
z^2 (1-u-z)^2 F_1^2 
+z^4 F_2^2
+u^4 F_3^2
\right].
\end{align}
Electroweak input parameters we use are
\begin{align}
G_F &= 1.1663788\cdot 10^{-5}/\text{GeV}^2,\quad \alpha = 1/137.036,\\
m_W &= 80.37\text{ GeV}, \quad m_Z = 91.19\text{ GeV},\\
\cos \theta_W &= m_W/m_Z, \quad m_e = 0.511\text{ MeV},
\end{align}
After numerical integration, we get the following decay rate,
\begin{equation}
    \Gamma(\pggg)  = \resultWZ.
    \label{eq:decayrate}
\end{equation}
The largest contribution to this number comes from the square of the sum of diagrams containing the $Z$-boson, followed by their interference with $W$-diagrams. Since the $Z$-diagrams are proportional to the small vector coupling of $Z$ to electrons, $\sim 1-4\sin^2\theta_W$, our result is sensitive to the assumed value of $\theta_W$. Higher-order electroweak radiative corrections tend to further decrease that vector coupling. A more careful discussion could mirror the analysis in \cite{Czarnecki:1995fw} where another low-energy, parity-violating process was considered: M{\o}ller scattering. 

Given the extreme smallness of our result, high accuracy is not warranted. Our result \eqref{eq:decayrate} can be treated as an upper limit on $\Gamma(\pggg)$, and likely exceeds the exact Standard Model prediction by about 50 per cent.  

The branching ratio of the three-photon channel is obtained by dividing Eq.~\eqref{eq:decayrate} by the width of the leading, two-photon decay, $\Gamma(\pgg)\simeq m_e\alpha^5/2$,
\begin{equation}
    \text{BR}(\text{p-Ps}\to 3\gamma) = 2.1\cdot 10^{-75}.
    \label{eq:BR}
\end{equation}
This branching ratio exceeds the partial result \cite{Pokraka:2017ore}, based on $W$-contributions only, by a factor of almost 50, despite the suppression of effects of the $Z$  boson by its small vector coupling. This is to a large extent caused by the presence of two widely separated mass scales in $Z$-diagrams and their resulting enhancement by the factor of $\ln(m_Z^2/m_e^2) \simeq 24$. 

Despite this enhancement, the branching ratio is smaller than estimated in Ref.~\cite{Bernreuther:1981ah} by a factor of about $5\cdot 10^{47}$. In that work, the branching ratio of $\pggg$ was estimated to be
\begin{align}    
    \text{BR}_{\text{Ref.~\cite{Bernreuther:1981ah}}}(\pggg) &=
     \alpha \left[ G_F m_e^2 (1-4\sin^2\theta_W)\right]^2\\ 
     & \simeq 10^{-27}.
     \label{eq:BN}
\end{align}
This estimate was based on the expectation that diagrams of the type shown in Fig.~\ref{fig:fermionloops}(c) contribute. We proved in Section \ref{sec: fermion loops} that they in fact vanish.

Finally, we present an analytical formula, based on the total $W$ contribution but only the logarithmically-enhanced part of the $Z$ part,
\begin{align}
\text{BR}(\pggg) &\simeq
\frac
{\alpha  G_F^2 m_e^{12} }  {m_W^8} 
 \frac{14700 \pi^2-145081}
   {1\,428\,840\,000\, \pi ^5 }
   \left[7+ 30 \cos^4\theta_W
   \left(1 - 4 \sin^2\theta_W\right) \ln
   \frac{m_Z^2}{m_e^2}
   \right]^2
   \\
   & = 2.6\cdot 10^{-75}.
   \label{eq:approx}
\end{align}
It exceeds the numerical value in Eq.~\eqref{eq:BR} by about a quarter.
The second term in the bracket (logarithmically enhanced $Z$-contribution) is about 6.7 times larger than the first (the $W$ contribution). This demonstrates the importance of $Z$ diagrams. One can also see that the interference between $Z$ and $W$ contributions is constructive. 

Comparing our  approximate analytical formula \eqref{eq:approx} with \eqref{eq:BN}, we see that Ref.~\cite{Bernreuther:1981ah} missed the suppression factor $(m_e/m_W)^8\simeq 10^{-42}$ as well as a significant dimensionless factor.

\section{Discussion and conclusion \label{sec:Conclusion}}

In this paper, we have completed the Standard Model prediction of the $\pggg$ decay rate by determining $Z$-boson contributions. The $Z$-boson mediated amplitude includes an extra scale $m_e$ in the loop as compared to the $W$-boson mediated amplitude. We handled that additional scale by expanding the loop integral in two regions, characterized by the loop momentum of order $m_Z$ and of order $m_e$, respectively. While it was suspected in~\cite{Pokraka:2017ore} that the large distance scale related to $m_e$ may enhance the $Z$-contribution relative to the $W$-contribution, our calculation shows, surprisingly, that the leading  amplitude is at $1/m_Z^6$ order, as shown in Eq.~\eqref{eq:F1} and Eq.~\eqref{eq:F4}. 

This is a far stronger suppression than estimated in~\cite{Bernreuther:1981ah}, where  the amplitude was thought to be of the $1/m_Z^2$ order. As a consequence, combining with the contribution from the $W$-boson mediated diagrams, the decay rate we find in Eq.~\eqref{eq:decayrate} is 47 orders of magnitude smaller than estimated in~\cite{Bernreuther:1981ah}. Further, we find that several diagrams, some of which were anticipated in~\cite{Bernreuther:1981ah} to contribute, turn out to vanish for various reasons, as discussed in Sec.~\ref{sec: fermion loops}.

Finally, we comment on the $\pi^0\to3\gamma$ amplitude. $\pi^0$ has $P=-1$, $C=+1$, and spin 0,  the same as $\pPs$. Because of these similarities, the $\pi^0\to3\gamma$ decay also requires mediation of the $W$ or $Z$ bosons as discussed in~\cite{Dicus:1975cz}. Comparing with the $\pggg$ decay, there are a few differences, as the hadronic effect inside the pion is important and there are extra mass scales in the $W$-boson mediated diagrams, due to all quarks being massive.  Therefore, the amplitude we get is not directly generalizable to  $\pi^0\to3\gamma$. However, given the similarity in the mechanism  generating both decays, it is still highly possible that the underlying reasons that led to the $1/m_{W,Z}^6$ behavior in $\pggg$ also apply to the pion decay. Therefore, a simple dimensional analysis as done in~\cite{Dicus:1975cz} might be insufficient for this decay, and it would be worthwhile to fully compute the $\pi^0\to3\gamma$ amplitude. 

\section*{Acknowledgments}

This research was supported by the Natural Sciences and Engineering Research 
Council of Canada (NSERC). RT was supported by an Undergraduate Research Scholarship of the Canadian Institute of Nuclear Physics.
We thank Guangpeng Zhang for collaboration in the early stage of this project.
\texttt{FORM}~\cite{Kuipers:2012rf},
\texttt{FeynCalc}~\cite{Shtabovenko:2025lxq,Shtabovenko:2016sxi,Shtabovenko:2020gxv,Mertig:1990an},
and \texttt{FIRE}~\cite{Smirnov:2019qkx} were used to evaluate the
amplitude. Figures were drawn with Asymptote \cite{Asymptote}.




\end{document}